\documentstyle [prl,aps,twocolumn]{revtex}
\begin{document}
\draft
\title{Reply to `Singularities of the mixed state phase'}
\author{ Jeeva Anandan$^{(1)}$, Erik Sj\"{o}qvist$^{(2)}$,
Arun K. Pati$^{(3)}$, Artur Ekert$^{(4)}$,
Marie Ericsson$^{(2)}$, Daniel K.L. Oi$^{(4)}$, and
Vlatko Vedral$^{(4)}$}
\address{$^{(1)}$ Department of Physics and Astronomy,
University of South Carolina, Columbia, SC 29208}
\address{$^{(2)}$ Department of Quantum Chemistry,
Uppsala University, Box 518, Se-751 20 Sweden}
\address{$^{(3)}$ Theoretical Physics Division, 5th Floor,
C. C., BARC, Mumbai-400085, India}
\address{$^{(4)}$ Centre for Quantum Computation, University
of Oxford, Clarendon Laboratory, Parks Road, Oxford OX1 3PU,
UK}
\maketitle

We agree with Bhandari \cite{bhandari01} that our mixed state
phase $\phi = \arg Tr(U_i\rho_0)$ is undefined in the special
cases
\begin{equation}
Tr(U_i\rho_0) =0. \label{singularity}
\end{equation}
However, for the example in our paper \cite{sjoqvist00} that
Bhandari criticizes $Tr(U_i\rho_0)=-1 \ne 0$. In this example of
interferometry with unpolarized neutrons, where one beam is
given a rotation of $2\pi$ radians, our mixed state phase shift
is $\pi$ (modulo $2\pi$), in agreement with the experiments.
But Bhandari claims that this phase shift is ``indeterminate''
because it could be $\pi$ or $-\pi$; but these two phases differ by $2\pi$.
So, the only difference between Bhandari's viewpoint
and ours is that
our phase is defined modulo $2\pi$, whereas Bhandari
argues that two phases that differ by $2\pi n$, $n$
integer, may be distinguished experimentally in a
history-dependent manner.

Bhandari's singularities are defined by (\ref{singularity}) in
relation to the input state $\rho_0$ and his non-modular phase is
associated with the evolution path that originates at $\rho_0$.
This phase has the disadvantage that it becomes undefined even at
points of the parameter space for which $Tr(U_i\rho_0) \ne 0$ if
the path has passed through a ``singularity'' (see Fig. 1 of
\cite{bhandari01}). But our phase modulo $2\pi$ is well defined at
all such points, as in the above example, because it does not
depend on the path. For the special case of spin 1/2 or qubit pure
state $\rho_0=|\psi><\psi|$, the singularity is the point opposite
to $\rho_0$ in the Bloch sphere or the Poincare sphere. The Pancharatnam phase is undefined for this pair of orthogonal states, which is not a problem
for this phase, and similarly (\ref{singularity}) is not a problem for our phase. The
interesting fringe shift in the interference pattern that Bhandari
obtains in his experiments (Refs. 3-5 of \cite{bhandari01}) when
the path goes around a singularity, but not around any other
point, may be explained by the change in $e^{i\phi}$, in which the
phase is defined modulo $2\pi$, instead of using his non-modular
phase $\phi$. Also, for arbitrary quantum systems in pure or mixed
states, these singularities may be detected, without the use of
the non-modular phase, by the vanishing of the visibility.

For arbitrary spin also Bhandari's approach does not give any
additional information as implied at the end of his Comment. The
geometric phases in this case may be obtained by parallel
transporting around the circuit $C$ traced by the direction of the
evolving spin quantization axis on the sphere $SU(2)/U(1)$. This
holonomy transformation gives \cite{an1987} the geometric phases
for the states with spin quantum numbers $j$ as
\begin{equation}
\beta_j = j\alpha \  ({\text{mod}} 2\pi), \ j=-J, -J+1,...,J
\label{modularphase}
\end{equation}
where $\alpha$ is the solid angle of either of the complementary
surfaces $S_1$ and $S_2$ on this sphere spanned by $C$, and the
spin $J$ is an integer or half-integer. The freedom to choose
either $S_1$ or $S_2$ {\it requires that the phase should be
defined modulo $2\pi$}, because their solid angles add to $4\pi$.
This is an interesting aspect of the Dirac monopole geometry which
gives rise to the geometric phases. The mixed state geometric
phase is then $\beta = \arg\{\sum_j \lambda_j \exp(i\beta_j)\}$,
where $\lambda_j$ are non degenerate eigenvalues of the density
matrix $\rho_0$ ($\lambda_j \geq 0, \sum_j \lambda_j =1$). Now,
(\ref{modularphase}) is equivalent to $\beta_j = j \alpha + 2\pi n
$ where $n$ is a particular integer. Suppose $C$ is infinitesimal.
Since $\beta_j$ is obtained in any experiment from $e^{i\beta_j}$,
we may instead regard its values corresponding to all possible
values of $n$ to be equivalent. In particular, both $\alpha$ and
$\beta_j$ may be chosen to be infinitesimal, which corresponds to
$n=0$. Then the spin quantum number $j=\beta_j/\alpha$ is obtained
from the known values of $\beta_j$ and $\alpha$, without having to
go around a ``singularity''.

In the geometrically analogous magnetic monopole case, this
corresponds to determining the magnetic charge by simply measuring
the field strength at the infinitesimal circuit. Bhandari's
history-dependent, non-modular phase implicitly chooses a gauge
that has the analog of a Dirac string whose intersection with the
sphere is his ``singularity.''  His phase then is defined using
the solid angle of one of the two surfaces $S_1,S_2$ that has no
singularity. But this is contained as a special case, with
appropriate choice of $n$, of the above more general treatment
that is valid in all gauges.


\begin{references}
\bibitem{bhandari01} R. Bhandari,
preceding comment.
\bibitem{sjoqvist00} E. Sj\"{o}qvist, A.K. Pati, A. Ekert,
J.S. Anandan, M. Ericsson, D.K.L. Oi, and V. Vedral,
Phys. Rev. Lett. {\bf 85}, 2845 (2000).
\bibitem{an1987}
J. Anandan and L.Stodolsky, Phys. Rev. D {\bf 35,} 2597 (1987);
J. Anandan, Phys. Lett. A, {\bf 129,} 201 (1988). These two
papers should have been the reference to the parallel transport
conditions (13) in \cite{sjoqvist00}, and not [19] as mistakenly
stated in \cite{sjoqvist00}.
\end{references}
\end{document}